\documentclass[prl,twocolumn,showpacs,showkeys,preprintnumbers,floatfix,footinbib,superscriptaddress]{revtex4-1}
\usepackage{amsfonts} 
\usepackage{amssymb} 
\usepackage{amsmath} 
\usepackage{graphicx} 
\usepackage{subfigure} 
\usepackage{array} 
\usepackage{dcolumn} 
\usepackage{bm} 
\usepackage{latexsym} 
\usepackage{longtable} 
\usepackage{hyperref} 
\usepackage{bbold}
\usepackage{color}
\usepackage{multirow}

\graphicspath{{./figs/}}

\newcommand{\qslash}[1]{\text{$\not \! #1$}}

\newcommand{\MeV}{\mathop{\rm MeV}\nolimits}
\newcommand{\GeV}{\mathop{\rm GeV}\nolimits}

\newcommand{\edited}[1]{{\color{black}{#1}}}

\begin{document}


\title{Neutron Electric Dipole Moment and Tensor Charges from Lattice QCD}
\author{Tanmoy Bhattacharya}
\email{tanmoy@lanl.gov}
\affiliation{Los Alamos National Laboratory, Theoretical Division T-2, Los Alamos, NM 87545}

\author{Vincenzo Cirigliano}
\email{cirigliano@lanl.gov}
\affiliation{Los Alamos National Laboratory, Theoretical Division T-2, Los Alamos, NM 87545}


\author{Rajan Gupta}
\email{rajan@lanl.gov}
\affiliation{Los Alamos National Laboratory, Theoretical Division T-2, Los Alamos, NM 87545}


\author{Huey-Wen Lin}
\email{hueywenlin@lbl.gov}
\affiliation{Physics Department, University of California, Berkeley, CA 94720}

\author{Boram Yoon}
\email{boram@lanl.gov}
\affiliation{Los Alamos National Laboratory, Theoretical Division T-2, Los Alamos, NM 87545}

\collaboration{Precision Neutron-Decay Matrix Elements (PNDME) Collaboration}
\preprint{LA-UR-15-24210}
%
\pacs{11.15.Ha, 
      12.38.Gc  
}
\keywords{neutron tensor charges, Lattice QCD, neutron EDM, split SUSY}
\date{\today}
\begin{abstract}
We present lattice QCD results on the neutron tensor charges
including, for the first time, a simultaneous extrapolation in the
lattice spacing, volume, and light quark masses to the physical point
in the continuum limit.  We find that the ``disconnected''
contribution is smaller than the statistical error in the ``connected''
contribution. Our estimates in the $\overline{\text{MS}}$ scheme at
$2\GeV$, including all systematics, are $g_T^{d-u}=1.020(76)$,
$g_T^d = 0.774(66)$, $g_T^u = - 0.233(28)$, and $g_T^s = 0.008(9)$.
The flavor diagonal charges determine the size of the neutron electric
dipole moment (EDM) induced by quark EDMs that are generated in many
new scenarios of CP-violation beyond the Standard Model (BSM).  We use
our results to derive model-independent bounds on the EDMs of light
quarks and update the EDM phenomenology in split supersymmetry with
gaugino mass unification, finding a stringent upper bound of $d_n < 4
\times 10^{-28} \, e$ cm for the neutron EDM in this scenario.
\end{abstract}
\maketitle
%
%
%

Low-energy precision measurements of neutron properties provide unique
probes of new physics at the TeV scale.  Searches for the neutron
permanent EDM \edited{$d_n$} have high sensitivity to new beyond
the standard model (BSM) CP-violating interactions.  Similarly, precision studies of
correlations in neutron decay are sensitive to possible BSM scalar and
tensor interactions.  To fully realize the potential of the vibrant
existing experimental neutron physics program~\cite{Dubbers:2011ns},
one needs to accurately calculate matrix elements of appropriate
low-energy effective operators within neutron states.  In this paper
we describe lattice QCD calculations of the neutron tensor
charges. \edited{In the future, these charges will be extracted with
  competitive precision from various measurements of the quark transversity
  distributions at
  JLab~\cite{Courtoy:2015haa}, and provide robust tests of the
  lattice results.  }

The flavor diagonal charges $g_T^{u,d,s}$ are needed to quantify the
contribution of the quark EDM to the neutron EDM and thus set bounds
on BSM sources of CP violation.  We find that the contribution of the
``disconnected'' diagrams to $g_T^{u}$, $g_T^{d}$ and $g_T^{s}$ are
small. Our results on these charges allow us to constrain split supersymmetry
models. \looseness-1


The isovector charge
$g_T^{d-u}$ is needed in the analysis of precision neutron $\beta$-decay. 
In Ref.~\cite{Bhattacharya:2011qm} we showed that to complement
experimental measurements of the helicity flip contributions to
neutron $\beta$-decay at the precision of planned experiments
($10^{-3}$ level), we need to calculate the iso-vector scalar and
tensor charges, $g_S^{d-u}$ and $g_T^{d-u}$, to about
$10\%$ accuracy. Results for $g_T^{d-u}$ presented here meet the desired
accuracy with control over all systematic errors, while $g_S^{d-u}$
requires $O(10)$ more statistics. 


Details of the lattice QCD calculations are given in a companion
paper~\cite{Bhattacharya:2015wna}.  Here we summarize the main points
and focus on the results using nine ensembles of $N_f=2+1+1$ flavors
of highly improved staggered quarks (HISQ)~\cite{Follana:2006rc}
generated by the MILC Collaboration~\cite{Bazavov:2012xda} and
described in Table~\ref{tab:ens}.  On these ensembles, we construct
correlation functions using Wilson-clover fermions, as these preserve
the continuum spin structure.  To reduce short-distance noise, all
lattices were ``HYP'' smeared~\cite{Hasenfratz:2001hp}. Extensive
tests were carried out on these nine HYP smeared ensembles to look for
the presence of exceptional
configurations~\cite{Bhattacharya:2013ehc}, a possible problem with
this mixed-action, clover-on-HISQ, approach. None were
detected. Issues of statistics, excited state contamination, operator
renormalization, lattice volume, lattice spacing and the chiral
behavior are detailed in~\cite{Bhattacharya:2015wna}.

\begin{table*}[th]
\begin{center}
\begin{ruledtabular}
\begin{tabular}{lc<{ }|ccccc|ccc}
Ensemble ID &   & $a$ (fm) & $M_\pi^{\text{sea}}$ (MeV) & $M_\pi~(\MeV)$ & $L^3\times T$    & $M_\pi L$ & $t_\text{sep}/a$ & $N_\text{conf}$ & $N_\text{meas}$  \\
\hline
a12m310 & \includegraphics[viewport=16 13 22 19]{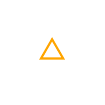}  & 0.1207(11) & 305.3(4) & 310.2(2.8) & $24^3\times 64$ & 4.55 & $\{8,9,10,11,12\}$& 1013 & 8104  \\
a12m220S& \includegraphics[viewport=16 13 22 19]{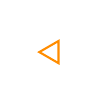}& 0.1202(12) & 218.1(4) & 225.0(2.3) & $24^3\times 64$ & 3.29 & $\{8, 10, 12\}$   & 1000 & 24000 \\
a12m220 & \includegraphics[viewport=16 13 22 19]{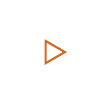}  & 0.1184(10) & 216.9(2) & 227.9(1.9) & $32^3\times 64$ & 4.38 & $\{8, 10, 12\}$   & 958  & 7664  \\
a12m220L& \includegraphics[viewport=16 13 22 19]{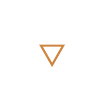}& 0.1189(09) & 217.0(2) & 227.6(1.7) & $40^3\times 64$ & 5.49 & $10$              & 1010 & 8080  \\
\hline                                                                                                                                                    
a09m310 & \includegraphics[viewport=16 13 22 19]{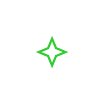}  & 0.0888(08) & 312.7(6) & 313.0(2.8) & $32^3\times 96$ & 4.51 & $\{10,12,14\}$    & 881  & 7048  \\
a09m220 & \includegraphics[viewport=16 13 22 19]{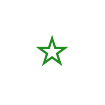}  & 0.0872(07) & 220.3(2) & 225.9(1.8) & $48^3\times 96$ & 4.79 & $\{10,12,14\}$    & 890  & 7120  \\
a09m130 & \includegraphics[viewport=16 13 22 19]{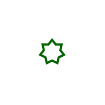}  & 0.0871(06) & 128.2(1) & 138.1(1.0) & $64^3\times 96$ & 3.90 & $\{10,12,14\}$    & 883  & 7064  \\
\hline                                                                                                                                                    
a06m310 & \includegraphics[viewport=16 13 22 19]{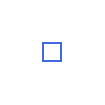}  & 0.0582(04) & 319.3(5) & 319.6(2.2) & $48^3\times 144$& 4.52 & $\{16,20,22,24\}$ & 1000 & 8000  \\
a06m220 & \includegraphics[viewport=16 13 22 19]{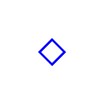}  & 0.0578(04) & 229.2(4) & 235.2(1.7) & $64^3\times 144$& 4.41 & $\{16,20,22,24\}$ & 650  & 2600  \\
\end{tabular}
\end{ruledtabular}
\caption{The parameters of the (2+1+1) flavor HISQ lattices are quoted
  from Ref.~\cite{Bazavov:2012xda}.  The symbols used in the plots are
  defined along with the ensemble ID.  All chiral analyses are carried out with
  respect to the clover valence pion masses $M_\pi$ which are tuned to 
  be close to the Goldstone HISQ pion masses $M_\pi^{\text{sea}}$. 
  We also give 
  the source-sink separations ($t_\text{sep}/a$) simulated,
  configuration analyzed ($N_\text{conf}$) and the total number of
  measurements ($N_\text{meas}$) made. Finite volume analysis is done
  in terms of $M_\pi L$. \looseness-1}
\label{tab:ens}
\end{center}
\vspace {-10pt}
\end{table*}

The flavor diagonal neutron charges $g_\Gamma^q$ are defined by 
$
 \langle n(p, s) \vert \mathcal{O}_\Gamma^q \vert n(p, s) \rangle
 = g_\Gamma^q \bar{u}_s(p) \Gamma u_s(p) \,,
$
with $O_\Gamma^q = \bar{q} \Gamma q$ and the spinors satisfying
$
 \sum_s u_s(\mathbf{p}) \bar{u}_s(\mathbf{p})   = (\qslash{p} + m)\,.
$
The interpolating operator we use to create$/$annihilate the relativistically normalized neutron
state $\vert n(p, s) \rangle$ is 
%
$
 \chi(x) = \epsilon^{abc} \left[ {q_1^a}^T(x) C \gamma_5 
            \frac{1}{2} (1 + \gamma_4) q_2^b(x) \right] q_1^c(x)
$
with color indices $\{a, b, c\}$, charge conjugation matrix $C$, and
$q_1$, $q_2$ the two different flavors of light quark fields.

\begin{figure*}
  \subfigure{
    \includegraphics[width=0.33\linewidth]{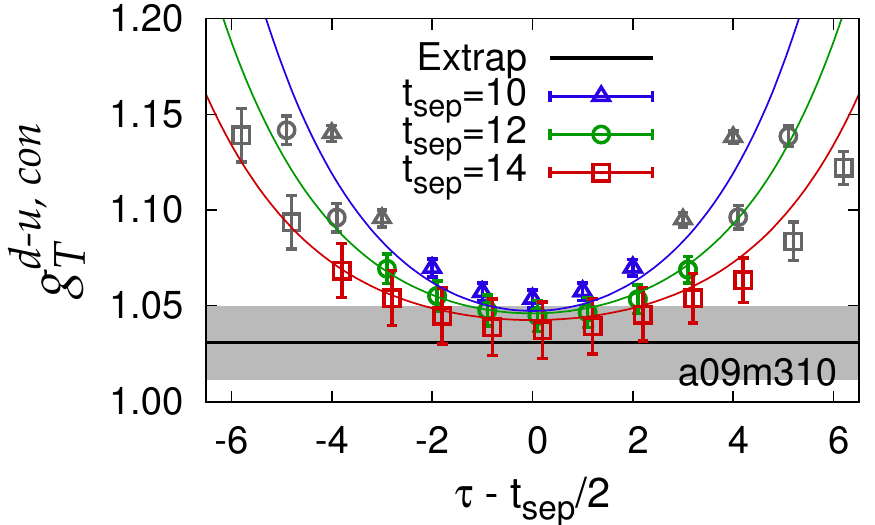}
    \includegraphics[width=0.33\linewidth]{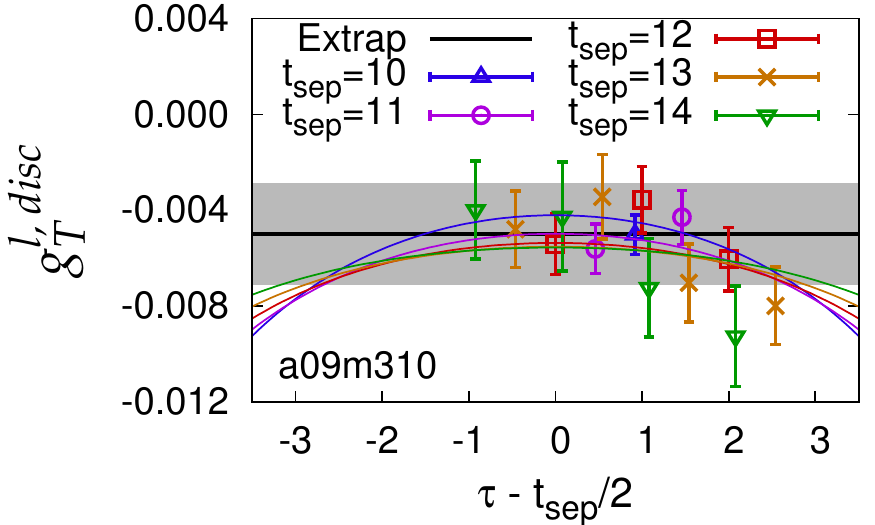}
    \includegraphics[width=0.33\linewidth]{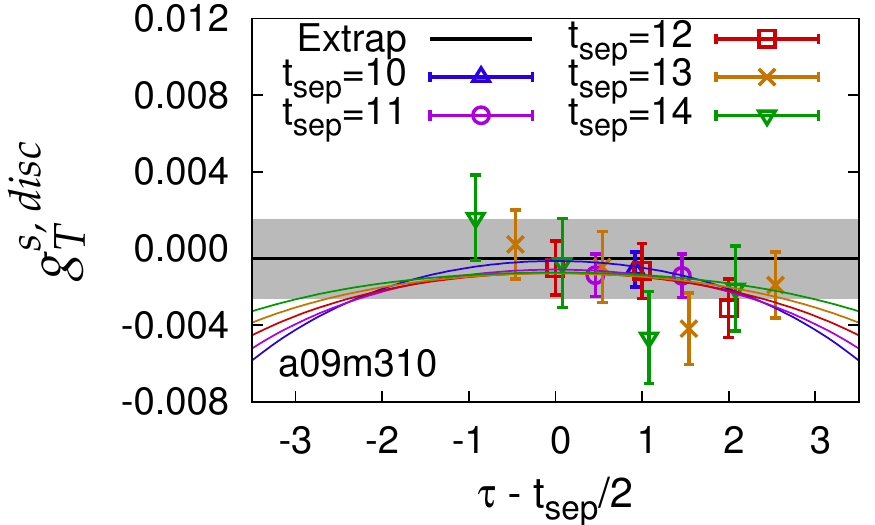}
}
\vspace {-16pt}
\caption{Fits illustrating the excited state contribution in the connected
  $g_T^{d-u}$ and light and strange disconnected diagram for the {\it
    a09m310} ensemble. The data points represent
  \(g_T^q(\tau,t_\text{sep})\) obtained from calculations at different source-sink separations
  $t_\text{sep}$ and operator
  insertion times $\tau$.  The solid black line and the gray band are
  the ground state estimate and error.
  \label{fig:excited}}
\end{figure*}

The zero-momentum projection of $\chi(x)$ couples to the ground state,
all radially excited states of the neutron, and multiparticle states.
To reduce the coupling to radially excited states we Gaussian smear
the quark fields in $\chi(x)$. To isolate the remaining excited state
contamination, we include two states in the analysis of the two- and
three-point functions at zero momentum~\cite{Bhattacharya:2015wna}.
Even though the excited state contribution is exponentially suppressed, we were able to
isolate the leading two unwanted matrix elements $\langle 0 |
\mathcal{O}_\Gamma | 1 \rangle$ and $\langle 1 | \mathcal{O}_\Gamma |
1 \rangle$, where $|0\rangle$ and $|1\rangle$ represent the ground and
first excited neutron states.  We find that the magnitude of $\langle
0 | \mathcal{O}_\Gamma | 1 \rangle$ is about $16\%$ of $\langle 0 |
\mathcal{O}_\Gamma | 0 \rangle$ and is determined with about $20\%$
uncertainty on all the ensembles, whereas $|\langle 1 |
\mathcal{O}_\Gamma | 1 \rangle| \sim \langle 0 | \mathcal{O}_\Gamma |
0 \rangle$, but has $O(100\%)$ errors.  As illustrated in
Fig.~\ref{fig:excited} for the {\it a09m310} ensemble, the overlap of
data in the center of the fit range for all the source-sink
separations $t_{\rm sep}$ indicates that excited state contamination
in the tensor charges is small and under control.  \looseness-1

The disconnected diagrams are estimated using a stochastic method
accelerated with a combination of the truncated solver method (TSM)
\cite{Collins:2007mh, Bali:2009hu}, the hopping parameter expansion
(HPE) \cite{Thron:1997iy, Michael:1999rs} and the all-mode-averaging
(AMA) technique~\cite{Blum:2012uh}.  In most cases, the disconnected
contribution is small and consistent with zero as illustrated in
Fig.~\ref{fig:excited} for the {\it a09m310} ensemble. This feature
was also observed in Ref.~\cite{Abdel-Rehim:2013wlz}. We find that the
light quark contribution is too noisy to extrapolate to the continuum
limit, so we do not include it in the central value.  We, however, use
the largest estimate, $0.0121$, on the coarsest ensemble {\it a12m310}
as an additional systematic error in $g_T^d$, $g_T^u$, and
$g_T^{d+u}$.

\begin{figure*}[th]
\includegraphics[width=0.95\linewidth]{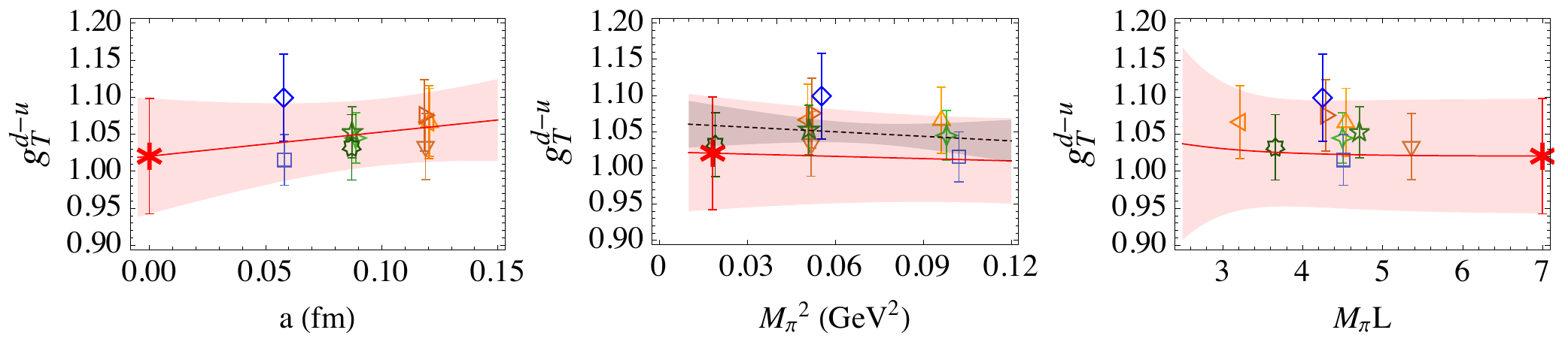}
\vspace {-12pt}
\caption{A simultaneous fit of neutron $g_T^{d-u}$ data versus $a$,
  $M_\pi^2$, and $M_\pi L$ using Eq.~\protect\eqref{eq:extrap}. The
  error band is shown as a function of each variable holding the other
  two at their physical value.  The data are shown projected on to
  each of the three planes.  The symbols are defined in
  Table~\protect\ref{tab:ens}.  The extrapolated value is marked by a
  red star.  The thin gray band and the dashed line within it in the middle
  panel show the fit versus $M_\pi^2$ assuming no dependence on the
  other two variables.
  \label{fig:chiral_gT_compare}}
\end{figure*}

\begin{figure}[thb]
\includegraphics[width=0.66 \linewidth]{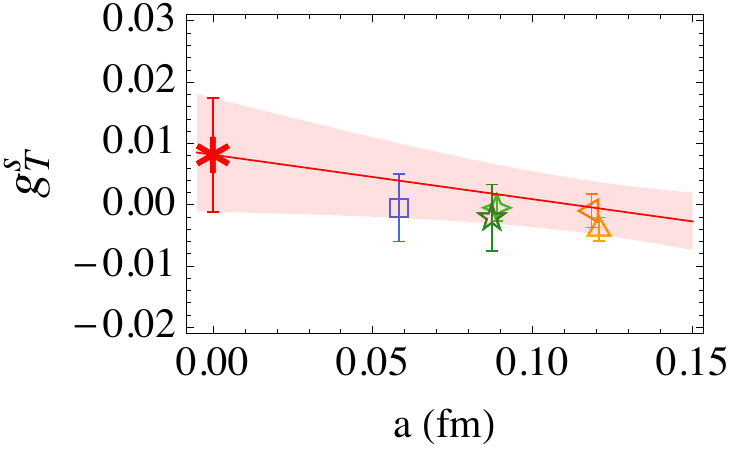} 
\vspace {-5pt}
\caption{The data for $g_T^s$ and intercept of the fit versus $a$ and
  $M_\pi$ on the $[g_T^{s}, a]$ plane. Notation is the same as in
  Fig.~\protect\ref{fig:chiral_gT_compare}.}
\label{fig:disc_s_extrap}
\vspace {5pt}
%
%
    \includegraphics[width=0.7\linewidth]{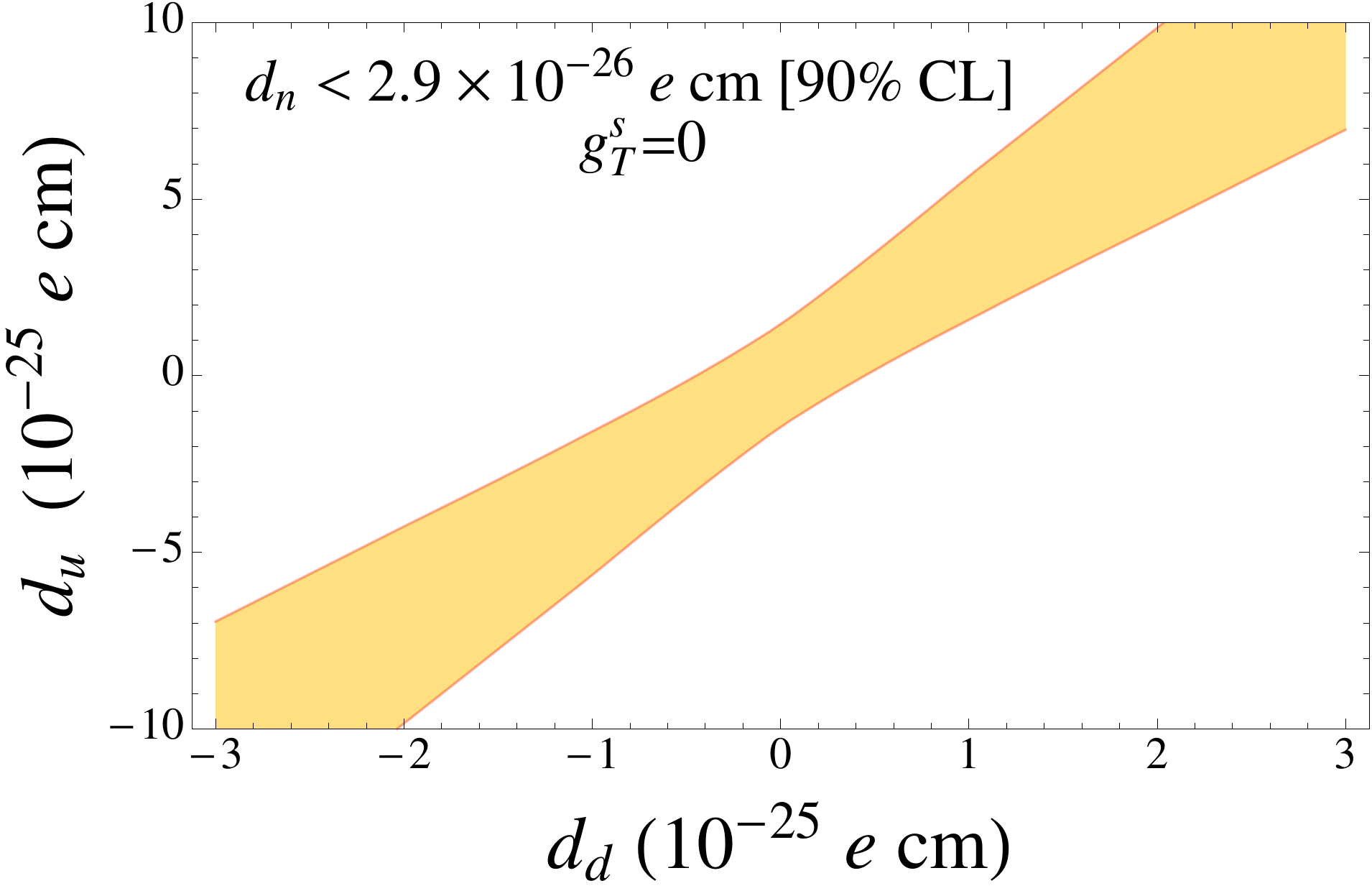}
\vspace {-5pt}
\caption{Bounds on $d_{u,d}$, \edited{defined in the  $\overline{\text{MS}}$ scheme at $2\GeV$,} with a $1\sigma$ slab prior 
  on $g_T^u$ and $g_T^d$ given in
  Eq.~\protect\eqref{eq:gTfinal} and $g_T^s=0$.
\label{fig:bounds}}
\vspace {-12pt}
\vspace{-5pt}
\end{figure}

The renormalization factor, calculated nonperturbatively in the
RI-sMOM scheme~\cite{Martinelli:1994ty,Sturm:2009kb} using the
iso-vector operator, contributes a significant fraction of the total
error.  \edited{The charges converted into the $\overline{\text{MS}}$
  scheme at $ 2\GeV$ are given in Table~\ref{tab:res-renorm} and
  Fig.~\ref{fig:chiral_gT_compare}. They are essentially
  flat in the three variables, lattice spacing $a$, the pion mass
  $M_\pi$ and the spatial lattice size $L$ .  We make a simultaneous fit to the
  data using the lowest order ansatz appropriate to our not fully
$O(a)$ improved clover-on-HISQ formulation: 
\begin{equation}
  g_T (a,M_\pi,L) = c_1 + c_2a + c_3 M_\pi^2 + c_4 e^{-M_\pi L} \,.
\label{eq:extrap}
\end{equation}
As discussed in~\cite{Bhattacharya:2015wna}, with current data the extrapolation to the
physical point ($M_\pi=135$~MeV, $a=0$, $M_\pi L = \infty$) is
insensitive to additional corrections.  The final renormalized
charges for the neutron~\footnote{In~\cite{Bhattacharya:2015wna} we
  follow the usual convention of defining the charges for the proton
  which are related to the neutron charges by the $u \leftrightarrow d$ interchange} are}
\begin{align}
g_T^{d\phantom{-u}}   & =  0.774(66) \, , \qquad g_T^{u\phantom{+u}}= -0.233(28) \,, \nonumber \\
g_T^{d-u} & =  1.020(76) \, , \qquad g_T^{d+u}  =  0.541(67) \,.
\label{eq:gTfinal}
\end{align}
The $\chi^2/\text{dof}$ for the fits are 0.1,  1.6, $0.4$ and $0.2$, 
respectively, with $\text{dof}=5$. Including the leading chiral
logarithms~\cite{deVries:2010ah} in Eq.~\eqref{eq:extrap}
gives similar results~\cite{Bhattacharya:2015wna}. 
$g_T^{s}$, after extrapolation
in the lattice spacing $a$ and $M_\pi^2$, is  
\begin{equation}
g_T^{s} = 0.008(9) \,,
\label{eq:gTs}
\end{equation}
with a $\chi^2/\text{dof} = 0.29$ with $\text{dof}=2$. The
intercept of the fit on the $[g_T^{s},a]$ plane is shown in
Fig.~\ref{fig:disc_s_extrap}. \looseness-1

Our result for $g_T^{d-u}$, with control over all systematic errors,
is in good agreement with other lattice
calculations~\cite{Syritsyn:2014saa,Constantinou:2014tga}.  The
LHPC~\cite{Green:2012ej} and RQCD~\cite{Bali:2014nma} Collaborations
also find no significant dependence on the lattice spacing and volume,
but do find a small dependence on the quark mass, so they extrapolate
only in the quark mass using linear/quadratic (LHPC) and linear (RQCD)
fits in $M_\pi^2$.  Their final estimates, $g_T^{d-u} = 1.038(11)(12)$
(LHPC) and $g_T^{d-u} = 1.005(17)(29)$ (RQCD) are consistent with ours.  A
fit to our data versus only $M_\pi^2$, shown as an overlay in
Fig.~\ref{fig:chiral_gT_compare} (center), gives a similarly accurate estimate
$g_T^{d-u} = 1.059(29)$ with a $\chi^2/{\rm dof} = 0.3$.


\begin{table*}[th]
\centering
\begin{ruledtabular}
\begin{tabular}{c|cccc|cc}
Ensemble ID & $g_T^{\text{con}, d}$ & $g_T^{\text{con}, u}$ & $g_T^{\text{con}, d-u}$ & $g_T^{\text{con}, d+u}$ & $g_T^{\text{disc}, l}$ & $g_T^{\text{disc}, s}$ \\
\hline
a12m310  & 0.852(37) & $-$0.215(12)  & 1.066(46) & 0.637(31) & $-$0.0121(23) & $-$0.0040(19) \\
a12m220S & 0.857(43) & $-$0.209(19)  & 1.066(50) & 0.649(44) & ---           & ---           \\
a12m220  & 0.860(40) & $-$0.215(15)  & 1.075(48) & 0.644(36) & $-$0.0037(40) & $-$0.0010(27) \\
a12m220L & 0.840(37) & $-$0.194(12)  & 1.033(45) & 0.647(33) & ---           & ---           \\ 
\hline
a09m310  & 0.840(28) & $-$0.2051(98) & 1.045(34) & 0.634(25) & $-$0.0050(22) & $-$0.0005(21) \\
a09m220  & 0.836(28) & $-$0.216(10)  & 1.053(34) & 0.619(25) & ---           & $-$0.0021(54) \\
a09m130  & 0.809(40) & $-$0.222(20)  & 1.032(44) & 0.587(45) & ---           & ---           \\
\hline
a06m310  & 0.815(29) & $-$0.199(10)  & 1.015(34) & 0.617(27) & $-$0.0037(65) & $-$0.0005(55) \\
a06m220  & 0.833(52) & $-$0.264(22)  & 1.099(59) & 0.569(55) & ---           & ---           \\
\end{tabular}
\end{ruledtabular}
\caption{Renormalized estimates of the connected ($g_T^{\rm con}$) and disconnected ($g_T^{\text{disc}}$) 
  contributions in the $\overline{\text{MS}}$ scheme at $2\GeV$. }
\label{tab:res-renorm}
\vspace {-2pt}
\end{table*}


%
%


%

Our results on the tensor charges have implications for the neutron EDM 
and CP-violation in BSM theories. 
At the hadronic scale, $\mu \sim O (1)$~GeV, \edited{after integrating
  out all heavy degrees of freedom} the dominant effect of new
CP-violating couplings in BSM theories is encoded in local operators
of dimension five and six. Leading, among them, are
the elementary fermion EDMs~\cite{Engel:2013lsa,Pospelov:2005pr}:

\begin{equation}
\delta {\cal L} _{\rm CPV}  \supset  
 \ - \  \frac{i e}{2}  \,  \sum_{f = u,d,s,e} \   d_f \  \bar{f}  \sigma_{\mu \nu} \gamma_5  F^{\mu \nu}     f  ~. 
\label{eq:Leff0}
\end{equation}
\edited{The contribution of the quark EDM $d_q$ to $d_n$ is~\cite{Ellis:1996dg,Bhattacharya:2012bf} }
\begin{equation}
d_n =  g_T^{u} \, d_u  \ + \ g_T^{d} \, d_d  \ + \ g_T^{s} \, d_s~,
\label{eq:dN}
\end{equation}
consequently, improved knowledge of $g_T^{q}$ combined with
experimental bounds on $d_n$ provides stringent constraints on new CP
violation encoded in $d_q$. 

\begin{figure}[th]
\includegraphics[width=0.9\linewidth]{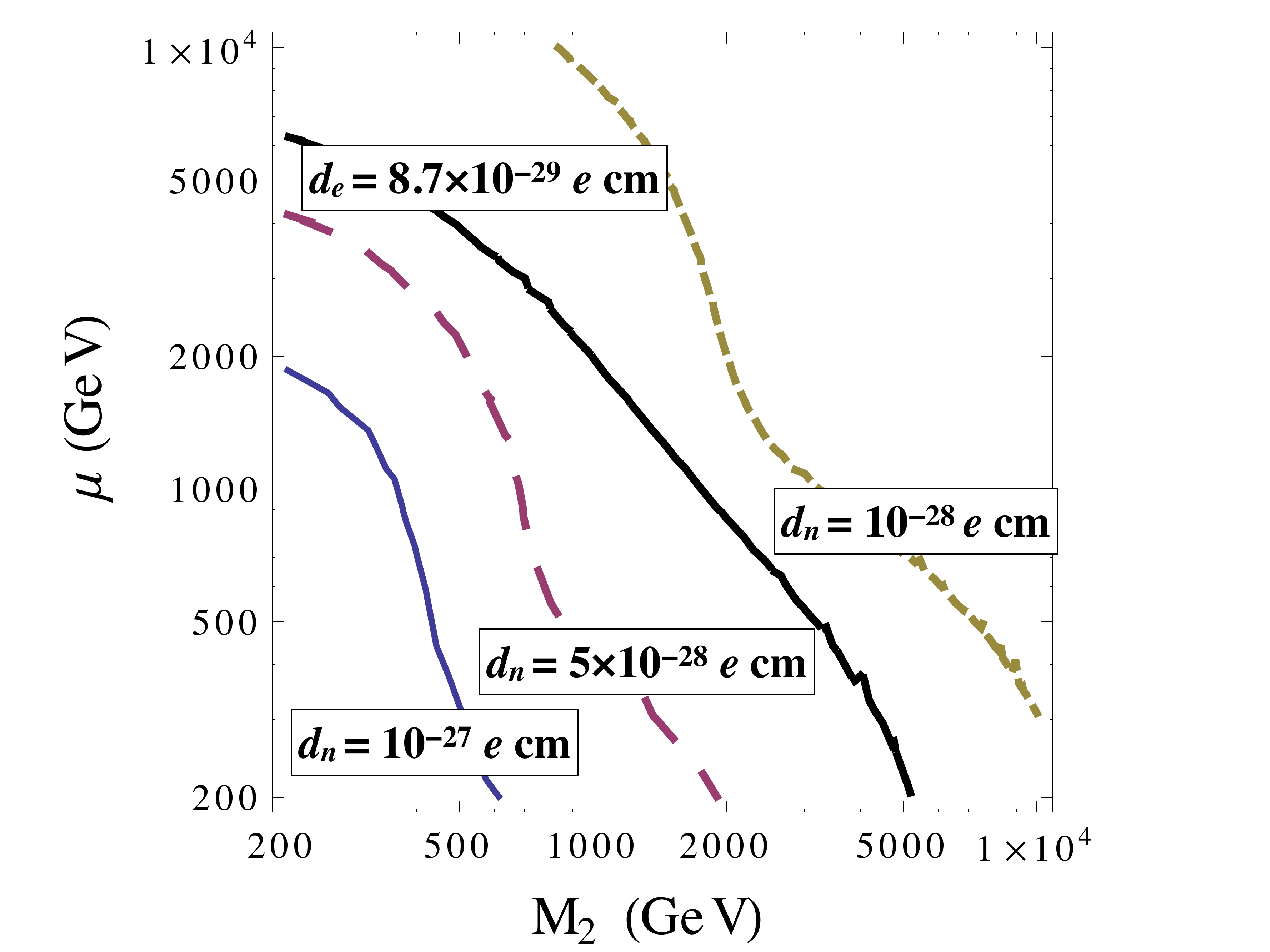}
\caption{Iso-level lines of $d_n$ and $d_e$ in split SUSY  in the $M_2$-$\mu$ plane using 
$\sin \phi = 1$, $\tan \beta = 1$, and central values of $g_T^{u,d,s}$. 
}
\label{fig:dnde}
\vspace {-12pt}
\end{figure}

Our calculation has the following  impact: 
(i) We reduce the uncertainty on $g_T^{u,d}$ from the $\sim 50\%$ of
previous QCD sum rules (QCDSR) estimates~\cite{Pospelov:2000bw} to the
10\% level. (For a comparison of the lattice results with the
Dyson-Schwinger~\cite{Pitschmann:2014jxa} and other
methods~\cite{Bacchetta:2012ty,Anselmino:2013vqa,Kang:2014zza,Kang:2015msa}
see~\cite{Bhattacharya:2015wna}.)
(ii) The central values of $g_T^{u,d}$ are roughly $3/5$ of the QCDSR and quark
model estimates~\cite{Pospelov:2000bw} widely used in phenomenological
studies of BSM CP violation.  
(iii) Bounding the strangeness tensor charge $g_T^{s}$ at the percent
level is important for a large class of models in which $d_q \propto
m_q$ since $m_s/m_d \sim 20$.  Our results imply that in such models
$g_T^{s} d_s$ may contribute up to $35\%$ of the total $d_n$ and the
current $O(1)$ fractional uncertainty in $g_T^{s} $ gives rise to the
largest uncertainty in $d_n$. \edited{The
  contribution of EDMs of heavier quarks to nEDM appears at
  two-loops and does not grow with $m_q$. In this work, we ignore 
  contributions of the charm (not calculated) and heavier quarks.}

%

While, in general, BSM theories generate additional CP-violating
operators in Eq.~(\ref{eq:Leff0}), there exist models in which the
fermion EDMs are the dominant sources of CP violation at low-energy,
thus controlling the pattern of hadronic and atomic EDMs.  For such
cases, using Eq.~(\ref{eq:dN}), our results on the tensor charges, and
the experimental limit on the neutron EDM~\cite{Baker:2006ts}, we show
90\% confidence level (CL) bounds on quark EDMs $d_{u,d}$ in
\edited{Fig.~\ref{fig:bounds}~\footnote{In deriving the bounds, we
    neglect a possible contribution of the strong CP violating phase
    $\Theta$. Such a contribution is absent in the Peccei-Quinn
    scenario~\cite{Peccei:1977hh}}}. \looseness-1

One notable scenario in which fermion EDM operators provide the
dominant BSM source of CP violation is ``split SUSY''~\cite{ArkaniHamed:2004fb,Giudice:2004tc,ArkaniHamed:2004yi},
in which all scalars, except for one Higgs doublet, are much heavier than the
electroweak scale.  This SUSY scenario achieves gauge coupling
unification, has a dark matter candidate, and avoids the most
stringent constraints associated with flavor and CP observables
mediated by one-loop diagrams involving scalar particles.
Contributions to fermion EDMs arise at two loops due to CP violating
phases in the gaugino-Higgsino sector, while all other operators are
highly suppressed~\cite{Giudice:2005rz,Li:2008kz}.  To illustrate the impact of
improved estimates of matrix elements in split SUSY, we use the analytic results
and setup of Ref.~\cite{Giudice:2005rz}, namely unified framework for
gaugino masses at the GUT scale and a scalar mass $\tilde{m} =
10^9$~GeV.  The light fermion EDMs $d_{e,u,d,s}$ depend on a single
phase $\phi$, on $\tan \beta$ [approximately through the
overall factor $\sin( \phi ) \sin (2 \beta)$] and on the gaugino
($M_2$) and Higgsino ($\mu$) mass parameters.  Following
Ref.~\cite{Giudice:2005rz} we set $\tan \beta = 1$, $\sin \phi = 1$
and present the results as contours in the $M_2$-$\mu$ plane, in the
range between $200$~GeV and $10$~TeV.
\begin{figure}[th]
\includegraphics[width=0.9\linewidth]{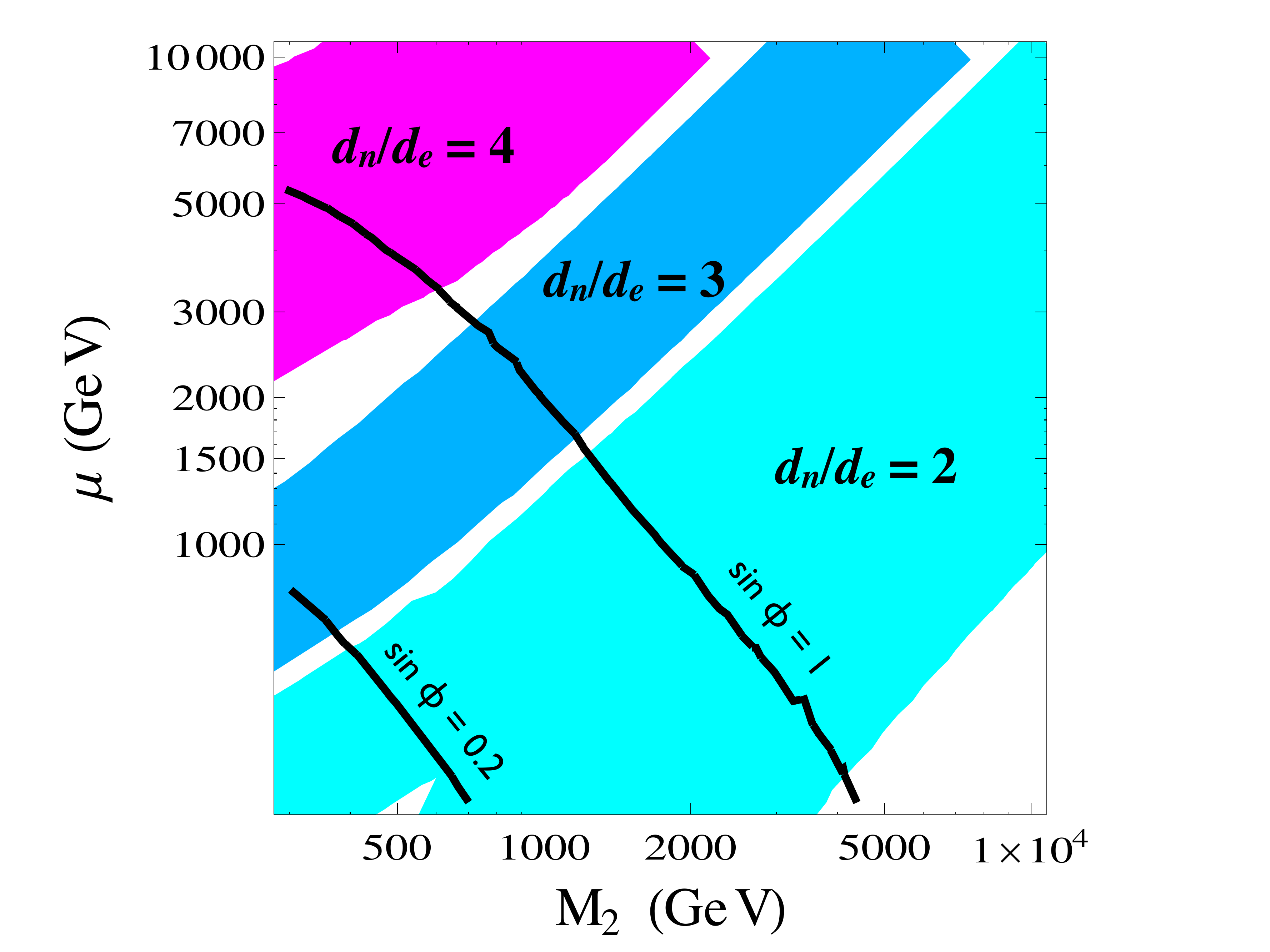}
\caption{ \edited{ Regions in $M_2$-$\mu$ plane corresponding to
    $d_n/d_e = 2, 3, 4$ in split SUSY, obtained by varying
    $g_T^{u,d,s}$ within our estimated uncertainties.  The lines
    correspond to $d_e = 8.7 \times 10^{-29} \, e$ cm for $\sin \phi 
    = 0.2, 1$.  }}
\label{fig:ratio}
\vspace {-12pt}
\end{figure}

Figure~\ref{fig:dnde} shows iso-level curves of $d_n$ as well as the
curve $d_e = 8.7 \times 10^{-29} \, e$ cm, corresponding to the
current 90\% C.L. limit~\cite{Baron:2013eja}.  For the neutron EDM we
use Eq.~(\ref{eq:dN}), evaluating both the $d_q$'s and the tensor
charges at the scale $\mu^{\overline{\rm MS}} = 2$~GeV.  Our result
for $d_n$ is appreciably smaller (factor of $\sim 3$) than the one in
Ref.~\cite{Giudice:2005rz}.  We have traced back this difference to
(i) our smaller values of the tensor charges compared to
QCDSR~\cite{Pospelov:2000bw}; (ii) different values for the light quark
masses: we use the PDG~\cite{Agashe:2014kda} central value $m_d
({\overline{\rm MS}}, \mu = 2 {\rm GeV}) = 4.75$~MeV, while the 
value corresponding to the quark condensate used in Ref.~\cite{Giudice:2005rz} 
is larger, $m_d ({\overline{\rm MS}}, \mu= 1 {\rm GeV} ) \approx 9$~MeV.\looseness-1

In Refs.~\cite{Giudice:2005rz, Abel:2005er}, it was pointed out that
the strong correlation between electron and neutron EDM would provide
a valuable experimental test of split SUSY.  To investigate this
further, in Fig.~\ref{fig:ratio} we present bands corresponding to
different values of $d_n/d_e$ in the $M_2$-$\mu$ plane, the thickness
of the bands reflects the behavior of $d_n/d_e [M_2, \mu]$ and the
uncertainty induced by the tensor charges, dominated by $g_T^{s}$.
The fact that we can draw disconnected bands for $d_n/d_e = 2,3,4$ is
a welcome consequence of our reduced uncertainties in $g_T^{q}$:
using the QCDSR input, each band would be as thick as the whole plot,
giving essentially no discrimination.

Finally, based on the current 90\% C.L. limit on $d_e$, we derive an
upper limit for the neutron EDM in split SUSY.  By maximizing the
ratio $d_n/d_e$ along the iso-level curves $d_e = 8.7 \times 10^{-29}
\, e$ cm corresponding to $\sin \phi \leq 1$, allowing
$g_T^{q}$ to vary in the lattice QCD ranges, we arrive at $d_n < 4
\times 10^{-28} \, e$ cm~\footnote{Note that with QCDSR matrix
  elements, the split-SUSY upper bound would be less stringent, namely
  $d_n < 1.2 \times 10^{-27} \, e$ cm.}.  Therefore,
observation of the neutron EDM between the current limit of $3 \times
10^{-26} \, e$ cm~\cite{Baker:2006ts} and $4 \times 10^{-28}
\, e$ cm would falsify the split-SUSY scenario with gaugino
mass unification.



\begin{acknowledgments}
We thank the MILC Collaboration for providing the 2+1+1 flavor HISQ
lattices used in our calculations.  Simulations were carried out on
computer facilities of (i) the USQCD Collaboration, which are funded
by the Office of Science of the U.S. Department of Energy, (ii) the
Extreme Science and Engineering Discovery Environment (XSEDE), which
is supported by National Science Foundation Grant No. ACI-1053575,
(iii) the National Energy Research Scientific Computing Center, a DOE
Office of Science User Facility supported by the Office of Science of
the U.S. Department of Energy under Contract No. DE-AC02-05CH11231; and 
(iv) Institutional Computing at Los Alamos National Lab. 
The calculations used the Chroma software
suite~\cite{Edwards:2004sx}. This material is based upon work supported by 
the U.S. Department of Energy, Office of Science of High Energy Physics under 
Contract No.~DE-KA-1401020 and the LANL LDRD program. The work of H.W.L. 
was supported by DOE Grant No.~DE-FG02-97ER4014. We thank 
Emanuele Mereghatti, Saul Cohen and Anosh Joseph for extensive discussions.
\end{acknowledgments}

%

%
\makeatletter
\ifx\@bibitemShut\undefined\let\@bibitemShut\relax\fi
\makeatother
\bibliography{ref} 

\end{document}